\documentclass{article}%
\usepackage{graphicx}
\usepackage{amsmath}%
\usepackage{amsfonts}%
\usepackage{amssymb}
\newtheorem{theorem}{Theorem}
\newtheorem{acknowledgement}[theorem]{Acknowledgement}

\begin{document}

\title{Modeling the propagation of a signal through a layered nanostructure:
\ Connections between the statistical properties of waves and random walks.}
\author{Gabriel A. Cwilich\\Yeshiva University\\Department of Physics\\500 W 185th Street\\New York, NY 10033, USA}
\maketitle

\begin{abstract}
It is possible to discuss the propagation of an electronic current through
certain layered nanostructures modeling them as a collection of random
one-dimensional interfaces, through which a coherent signal can be transmitted
or reflected while being scattered at each interface. \ We present a simple
model in which a persistent random walk ( the \textquotedblleft
t-r\textquotedblright\ model in 1-D) is used as a representation of the
propagation of a signal in a medium with such random interfaces.

In this model all the possible paths through the system leading to
transmission or reflection can be enumerated in an expansion in the number of
loops described by the path . This expansion allows us to conduct a
statistical analysis of the length of the paths for different geometries and
boundary conditions and understand their scaling with the size of the system.
By tuning the parameters of the model it is possible to interpolate smoothly
between the ballistic and the diffusive regimes of propagation. An extension
of this model to higher dimensions is presented. \ We show Monte Carlo
simulations that support the theoretical results obtained.

\end{abstract}

\section{Introduction}

The seminal work of Anderson\cite{Anderson} raising the possibility that
disorder can lead to non-diffusive behavior (the so called localized regime)
refocused the attention of the Physics community on the problem of the
propagation of waves in disordered systems. In the last two decades new
theoretical ideas (like the scaling theory of localization\cite{gang of four},
weak localization\cite{Lagendijk}, universal conductance
fluctuations\cite{Patrick Lee} and Wigner dwelling times\cite{Van Albada y
yo}) were advanced, and a new field (soon called Mesoscopic Physics) emerged.
It reached and influenced many experimental areas, among them electronic
systems\cite{Stone-Webb}, microwaves\cite{Genack}, optics\cite{Soukulis},
acoustics\cite{Fink}, geophysics\cite{Campillo}, laser physics\cite{Wiersma},
medical physics\cite{Wang} and atomic physics\cite{Kaiser}.

It has become an extremely important problem in this field to understand what
should be the signature of the propagation of a signal in the different
regimes (ballistic, diffusive, localized) since concomitant phenomena, like
absorption can complicate the interpretation of experimental
results\cite{Wiersma2} . It is for that reason that theoretical analyses of
the characteristics of the propagation, and in particular its statistical
properties\cite{Kogan}\cite{yo-OSA}, are of great interests, since those
properties have become recently experimentally accessible\cite{Genack}%
\cite{Sebbah}.

When the inelastic scattering length in a system is large compared to its
size, the wave propagates coherently in the sense of its phase being preserved
while its direction is randomized by elastic scattering processes with the
impurities constituting the random medium. It is natural, then, to establish
connections between the coherent propagation of a wave and the statistical
theory of random walks\cite{Feller}, where the continuum limit is known to
lead to diffusive theory\cite{weissrubin}. \ This connection is relevant to
the propagation of an electronic current through certain layered
nanostructures, since they can be\ modeled as a collection of random
one-dimensional interfaces, through which a coherent signal can be transmitted
or reflected while being scattered at each interface.

The connection between the effect of the disorder of the propagating medium
and the statistical randomness of the mathematical description of the
diffusion has not been satisfactorily clarified yet. One aspect that did not
receive enough attention in this approach is the fact that, since the
scattering with the impurities is not isotropic and the cross section is
normally enhanced in the forward direction, the statistical jumps are not
independent from each other. In other words the probability distribution of
each step of these persistent random walks (PRW) is dependent on the previous
step\cite{weiss}. \ The PRW are more difficult to study than the standard
random walks, where the probability distribution of each step is independent
of what happened in the previous steps. This can be seen from the fact that,
for example, while a standard random walk can be easily mapped into a chain of
non interacting spins (in the presence of a magnetic field, if the random walk
is biased), the same mapping for \ the PRW leads to a chain of
nearest-neighbor interacting spins (the full Ising model for the case of a 1-D
\ PRW)\cite{yo-OSA}. \ In particular in this work we will explore the
connection between the propagation of a wave and the 1-D version of the PRW
called the \textquotedblleft t-r\textquotedblright\ model\cite{martinez}, in
which a particle moves with probability $t$ in the same direction as in the
previous step and with probability $r$ reverses direction $(t+r=1)$. \ This
type of model is more suited than the standard random walk model to explore
the quasiballistic regime, which becomes more important in transport phenomena
in nanostructures.

\section{The model}

We will consider a model of the propagation of a wave first introduced for
numerical purposes by Edrei, Kaveh and Shapiro\cite{ed-kav-shap}, and later
applied extensively by Vanneste \textit{et al}\cite{seb2}\textit{\ }and
Sebbah\cite{seb3}. Each site is represented by a scattering matrix
$\mathbf{S}$ of dimension $2D\times2D$ \ ($D$ being the
dimensionality)\ connecting the $2D$ incoming amplitudes at time $t$, with the
$2D$ outgoing ones at time $t+1$. These amplitudes can be regarded as residing
in the bonds (2 joining each site to its nearest neighbor, one in each
direction). These bonds represent free propagation between the sites, and
their phase can be included in the $S$ matrix. The wave function at time $t$
is, thus, determined by the complex numbers associated with each of the bonds
at that time. Numerically this has the advantage of letting the stationary
wave being built step by step in time, from an initial input without solving a
huge diagonalization problem. The matrices $\mathbf{S}$ representing each
scatterer are unitary and symmetric reflecting, respectively, the energy
conservation and time reversal symmetry inherent to the problem. If we adopt
the simplifying assumption that the scatterers themselves are symmetric, these
matrices simplify even further, since the following relations can be easily
obtained for their elements
\begin{align}
r^{2}+t^{2}+2(D-1)d^{2}  & =1\label{cond-rtdfi}\\
r\,t\,\cos[\varphi_{r}-\varphi_{t}]+(D-1)d^{2}  & =0\nonumber\\
r\,d\,\cos[\varphi_{r}-\varphi_{d}]+t\,d\cos[\varphi_{t}-\varphi
_{d}]+(D-2)d^{2}  & =0\nonumber
\end{align}
where $\{r,\varphi_{r}\},\{t,\varphi_{t}\}$ and $\{d,\varphi_{d}\}$ are the
amplitudes and phases of the elements for reflection, transmission, and
turning (in 1D, $d=0$). The matrices $\mathbf{S}$ have, thus, only two (in 1D)
or three \ (in D%
$>$%
1) independent parameters. \ For example, in 1D the scattering matrix
representing one scatterer can be parameterized as:
\begin{equation}
\left[
\begin{array}
[c]{cc}%
{\large r} & {\large I\,t}\\
{\large I\,t} & {\large r}%
\end{array}
\right]  \exp{\large [I\,\varphi]\;\hspace{1in}(r}^{2}{\large +\;t}%
^{2})=1\label{1dmatrix}%
\end{equation}
In this simple version of the model the distances are measured in units of the
mean free path (the distance between scattering elements) and the phase
velocity of the wave is one. The effect of the disorder in this model has been
considered both analytically\cite{stepabra} and numerically\cite{andereck}, by
assuming a distribution of values for the variable $r$, linked to the
reflectivity of the scatterers. Alternatively, one can consider the effect of
disorder in the variable $\varphi$, more closely linked to the distance
between scatterers\cite{cwilklein}.

We will be interested in considering a signal incident from the left on a
system composed of $N$ such scatterers, and in monitoring what is reflected or
transmitted through it at different times, i.e. the output to the left of
scatterer $1$ and to the right of scatterer $N$, respectively. It is important
to observe that while calculating the response of the system to an applied
pulse implies adding all the paths of a certain length (that will interfere by
arriving simultaneously to the boundary), the problem of determining the
response to a continuous wave is simpler, since in the stationary state it is
sufficient to add \textbf{all} the transmitting or reflecting paths as long as
the phase factors associated with the propagation through each scatterer,
$\exp[I\,k]$, are included. $k$ represents here the wavevector of the incident wave.

\section{Loop expansion \ \ \label{loop}}

The amplitude factor for any particular path reaching the boundaries of the
system can be easily written, and we can sum all the possible paths by
performing a loop expansion. For example, for a system of \ two scatterers the
amplitudes associated with the simple paths illustrated in figure 1(a) are,
respectively, $(I\,t\exp[I\,k])^{2}$ , $(I\,t\exp[I\,k])^{2}(r\exp[I\,k])^{2}$
and $(I\,t\exp[I\,k])^{2}(r\exp[I\,k])^{4}$; the total amplitude for
transmission through the two scatterers, when all possible paths are
considered, becomes then:
\begin{equation}
t_{2}=\frac{(I\,t\exp[I\,k])^{2}}{1-(r\exp[I\,k])^{2}}\equiv\frac
{(I\,t\exp[I\,k])^{2}}{1-l_{1}}\label{t2wave}%
\end{equation}
For the three scatterers in figure 1(b), the amplitude factors associated with
the paths shown are $(I\,t\exp[I\,k])^{3}$, $(I\,t\exp[I\,k])^{3}%
(r\exp[I\,k])^{2}(I\,t\exp[I\,k])^{2}$ \ and $\ (I\,t\exp[I\,k])^{3}%
(r\exp[I\,k])^{2}(I\,t\exp[I\,k])^{2}$\ $(r\exp[I\,k])^{4}$, respectively. The
\textquotedblleft bare\textquotedblright\ loop of length 2 has an amplitude
factor of $\ l_{2}\equiv(r\exp[I\,k])^{2}(I\,t\exp[I\,k])^{2}$, and when fully
\textquotedblleft dressed\textquotedblright\ by all the possible loops of
length 1 becomes:%
\begin{figure}
[ptb]
\begin{center}
\includegraphics[
trim=0.816143pt 0.820785pt -0.816155pt -0.820786pt,
natheight=191.459106pt,
natwidth=469.317688pt,
height=142.6793pt,
width=347.5183pt
]%
{GQ6C7P00.wmf}%
\caption{(a) \ Three transmitting paths through a system of two scatterers,
showing no loops, one loop of length one, and two loops of length one,
respectively. \ (b) Three transmitting paths through a system of three
scatterers, showing no loops, one loop of length two, and the same loop of
length two \textquotedblleft dressed\textquotedblright\ by two loops of length
one, respectively.}%
\end{center}
\end{figure}
%

\[
\;\widetilde{l_{2}}=\frac{(r\exp[I\,k])^{2}(I\,t\exp[I\,k])^{2}}{\left(
1-(r\exp[I\,k])^{2}\right)  ^{2}}\equiv\frac{l_{2}}{(1-\widetilde{l}_{1})^{2}}%
\]
We see that the factor associated with a \textquotedblleft
bare\textquotedblright\ loop of length $n$\ is simply $l_{n}\equiv
(r\exp[I\,k])^{2}(I\,t\exp[I\,k])^{2(n-1)}$. In terms of it we can obtain a
recursive expression for the general \textquotedblleft
dressed\textquotedblright\ loop of order $n$
\[
\widetilde{l_{1}}=l_{1}\;\;\ ;\;\;\ \widetilde{l_{2}}=\frac{l_{2}%
}{(1-\widetilde{l}_{1})^{2}}\;\ \;\ ;\;\widetilde{l_{3}}=\frac{l_{3}%
}{(1-\widetilde{l_{1}})^{4}(1-\widetilde{l}_{2})^{2}}\;
\]

\begin{equation}
\;\widetilde{l_{n}}=\frac{l_{n}}{(1-\widetilde{l_{1}})^{2(n-1)}(1-\widetilde
{l_{2}})^{2(n-2)}.....(1-\widetilde{l}_{n-1})^{2}}\label{ldressedn}%
\end{equation}
It is possible now to add all the paths through a chain of $N$ scatterers, the
amplitude of the transmission through the system, and express it in terms of
the loops:%

\[
t_{1}=I\,t\,\exp[I\,k]\;\;;\;\;t_{2}=\frac{(I\,t\,\exp[I\,k])^{2}%
}{(1-\widetilde{l_{1}})}\;\;;\;\;t_{3}=\frac{(I\,t\,\exp[I\,k])^{3}%
}{(1-\widetilde{l_{1}})^{2}(1-\widetilde{l_{2}})}%
\]

\begin{equation}
\;t_{n}=\frac{(I\,t\,\exp[I\,k])^{n}}{(1-\widetilde{l_{1}})^{(n-1)}%
(1-\widetilde{l_{2}})^{(n-2)}.....(1-\widetilde{l}_{n-1})}\label{wavetn}%
\end{equation}
As expected, $\;\widetilde{l_{n}}$ is proportional to $t_{n}{}^{2}$. \ A very
similar expansion yields, for the reflection amplitudes:
\[
r_{1}=r\exp[I\,k]\;\;;\;\;r_{2}=r_{1}+\frac{(I\,t\exp[I\,k])^{2}%
\,(r\exp[I\,k])}{(1-\widetilde{l_{1}}\,)}%
\]%
\[
r_{3}=r_{2}\;+\;\frac{(I\,t\exp[I\,k])^{4}\,(r\exp[I\,k])}{(1-\widetilde
{l_{1}}\,)^{3}(1-\widetilde{l_{2}}\,)}%
\]

\begin{equation}
r_{n}=r_{n-1}\;+\;\frac{(I\,t\exp[I\,k])^{2(n-1)}(r\exp[I\,k])}{(1-\widetilde
{l_{1}})^{(2n-3)}(1-\widetilde{l_{2}})^{(2n-5)}.....(1-\widetilde{l}%
_{n-2})^{3}\,(1-\widetilde{l}_{n-1})}\label{wavern}%
\end{equation}
The expression for the amplitude of reflection through the system admits a
clear interpretation in terms of a sum of all the paths that penetrate up to a
certain depth in the system: $r_{1}$corresponds to the paths that do not enter
into the system at all, $r_{2}-r_{1}$ represents the paths that penetrate up
to a depth $1$, $r_{3}-r_{2}$ are the paths that penetrate up to a depth $2$, etc.

The connection of this expansion for the transmission and reflection
amplitudes of the propagating wave with the PRW is provided by the fact that
we can write exactly the same expansion in loops in the \textquotedblleft
t-r\textquotedblright\ model, for all the possible paths of a random walker
leading to transmission or reflection. Here transmission means the random
walker arriving to a point further to the right of position $N$, and
reflection the walker reaching a point to the left of position $1$. In
particular, if we ignore the phase factors $\exp[I\,k]$ and the imaginary unit
in front of $t$, the amplitude factors discussed above become the
probabilities of the corresponding PRW paths, and Equations (\ref{ldressedn}%
),(\ref{wavetn}) and (\ref{wavern}) describe the \textquotedblleft
dressed\textquotedblright\ loops, probability of transmission and probability
of reflection in the case of a PRW as well, since for a random walk the
\textbf{total} probability of transmission (reflection) is simply the sum of
the probabilities of each of the paths leading to transmission (reflection).
Of course in each case, for a PRW the connection between $t$ and $r$ is not
provided by (\ref{1dmatrix}), but by the simple expression $t+r=1$.

It is easy to show now that the probability of transmission for a PRW becomes, then:%

\begin{equation}
t_{n}=\frac{t}{1+(N-1)\,r}\label{tnmrw}%
\end{equation}
and the probability of reflection becomes simply $\ r_{n}=1-t_{n}$.
\ Identifying $N$ with the length of the system $L$, and taking into account
that in this model $<\cos(\vartheta)>=t-r$, we can see that in the limit of
large $N$ the transmission tends to the diffusive limit $t(L)=(l^{\ast}/L)$.

In the case of a wave, the transmission or reflection probabilities are given
by the modulus square of the expressions calculated in (\ref{wavetn}) and
(\ref{wavern}). This implies sums of the type:
\begin{equation}
\left\vert \sum_{i}A_{i}\right\vert ^{2}=\sum_{i}\left\vert A_{i}\right\vert
^{2}+\sum_{i\neq j}A_{i}A_{j}^{\ast}\label{splitsum}%
\end{equation}
It is easy to see that, when adding the transmitting paths%
\'{}
amplitudes, if we consider the contribution of only the first term in the
expression above (neglecting the correlations between the paths) we will
reobtain (\ref{tnmrw}), where the roles of $t$ and $r$ are played by the
transmission coefficient $t^{2}\equiv T$, and the reflection coefficient
$r^{2}\equiv R$. It is precisely the overlap between paths (the second term)
that leads to corrections to the diffusion theory results.

\section{Statistics of the paths}

In the case of the PRW, knowing the probabilities\ $p_{i\text{ }}$and the
lengths $l_{i\text{ }}$of all the paths both in transmission and reflection,
we are now in a position to study some of their statistical properties.
Defining a generating function $\ F_{tr}\,(s)=\sum\,p_{i}\,s^{l_{i}}$, where
we sum over all the transmitted paths, and a similar definition for
$F_{ref}\,(s)$ where the sum is over all reflected paths, we can calculate the
average length of the paths as%

\[
<l>\;=\;\lim\limits_{s\rightarrow1}\,\,\frac{F^{^{\prime}}(s)}{F(s)}\;,
\]
obtaining for the average length:
\begin{equation}
<l>_{tr}=\,N\;\frac{1+(N-2)\,r+\frac{1}{3}(N-2)(N-1)\,r^{2}}%
{[1+(N-1)\,r][1-r]}\label{lavertr}%
\end{equation}

\begin{equation}
<l>_{ref}=\,\frac{N+\frac{1}{3}(2N-1)(N-1)\,r}{1+(N-1)\,r}\label{laverref}%
\end{equation}
In the diffusive limit discussed above, these expressions reproduce the well
known results of diffusive theory\cite{weitz}: $<l>_{tr}\!(L)\rightarrow
(L^{2}/\,l^{\ast})$. Analogously, $<l>_{ref}(L)\rightarrow(2/3)L$, independent
of the mean free path. The expressions (\ref{lavertr}) and\ (\ref{laverref})
are exact, within the model, and remain valid in the diffusive as well as the
ballistic limit $r\rightarrow0$, and can be used to interpolate between them.
This can be of interest in any situation in which the ballistic paths play an
important role as in medical imaging.

In a similar way one can express in terms of \ F(s) the variance of the path
length distribution, in transmission or reflection%

\[
<(l-<l>)^{2}>\;=\;\lim\limits_{s\rightarrow1}\,\;\,\frac{F^{^{\prime\prime}%
}(s)+F^{^{\prime}}(s)}{F(s)}-[\frac{F^{^{\prime}}(s)}{F(s)}]^{2}%
\]
obtaining the expressions:%

\begin{equation}
var_{tr}=\frac{N(N^{2}-1)\,r^{2}}{3}\frac{[N+\frac{2}{5}(N-2)(2N-1)\,r+\frac
{2}{15}(N-1)(N-2)(N-3)\,r^{2}]}{(1-r)^{2}[1+(N-1)\,r]^{2}}\label{vartran}%
\end{equation}

\begin{equation}
var_{ref}=\frac{(N^{2}-1)\;\mathbf{P}(r)}{3\,(1-r)^{2}\,\,[1+(N-1)r]^{2}%
}\label{vrf}%
\end{equation}
$\mathbf{P}(r)$ is $1+(2N-3)r+\frac{2}{15}(14N^{2}-30N+19)r^{2}+\frac{4}%
{15}(2N-1)(N-1)(N-2)r^{3}$ here. These exact expressions, which again allow us
to extrapolate between the diffusive and ballistic regimes, in the diffusive
limit tend to $(L^{2}/\,l^{\ast})^{2}$ and $\ (L^{3}/l^{\star})$ respectively.
In the case of transmission the variance scales with the size of the system as
the square of the average length (non-self averaging property) while in the
case of reflection it scales even faster than it, exhibiting the well known
noisy character of reflection\cite{weitz}.

Figures 2, 3 and 4 exhibit the results of numerical\ simulations of a
persistent random walker on a system of size $N=100$.
\begin{figure}
[ptb]
\begin{center}
\includegraphics[
natheight=173.699203pt,
natwidth=225.478897pt,
height=181.9791pt,
width=235.7388pt
]%
{GQ6C7P01.wmf}%
\caption{Fraction of transmitted paths ($\cdot$) and of reflected paths (o)
for different values of the transmission parameter t in a numerical simulation
of up to 210,000 random walkers on a system of size $N=100$. The full lines
correspond to equation \ (\ref{tnmrw}) for $t_{100}$ , the probability of
transmission, and $r_{100}$ the probability of reflection. }%
\end{center}
\end{figure}
\begin{figure}
[ptbptb]
\begin{center}
\includegraphics[
natheight=199.798996pt,
natwidth=468.297699pt,
height=144.7793pt,
width=337.3784pt
]%
{GQ6C7P02.wmf}%
\caption{Average length of (a) the transmitted paths and (b) the reflected
paths for the same simulation. The full lines correspond to the equations
(\ref{lavertr}) and ( \ref{laverref}) for $N=100$.}%
\end{center}
\end{figure}
\begin{figure}
[ptbptbptb]
\begin{center}
\includegraphics[
natheight=165.119202pt,
natwidth=407.578003pt,
height=135.7193pt,
width=332.9384pt
]%
{GQ6C7P03.wmf}%
\caption{Standard deviation of the distribution of the length of (a) the
transmitted paths and (b) the reflected paths for the same simulation. The
full lines correspond to the square root of equation (\ref{vartran}) and
equation (\ref{vrf}) for $N=100$. }%
\end{center}
\end{figure}
The transmission and reflection probabilities, the average length of the
transmitted and reflected paths and the variance for the path lengths in
transmission and reflection are represented as a function of the transmission
parameter $t.$ In all cases the agreement with the expression \ref{tnmrw}
\ and (\ref{lavertr}) to (\ref{vrf}) is excellent.

An interesting problem is how to define the probabilistic weights to perform a
similar statistical analysis of the length of the paths in the case of a
\textit{wave}, since the amplitudes are complex numbers with a phase, and the
different paths interfere with each other. It makes sense to use the
\textquotedblleft probabilities\textquotedblright\ (in the sense of the moduli
of the complex amplitudes squared) as those probabilistic weights. As the
discussion at the end of section \ref{loop} shows, those weights describe the
diffusive behavior of the system. Adopting this ansatz we can define again the
generating function as in the case of the PRW, and we can reobtain all the
results in expressions (\ref{lavertr}) to (\ref{vrf}), again replacing $t$ and
$r$ by the transmission coefficient $T$ and the reflection coefficient $R$, respectively.

\section{Higher dimensions}

All this analysis and the expansion in terms of loops can be easily
generalized to a system in a higher number of dimensions. To fix ideas let us
consider, instead of a row of $N$ scatterers, a strip of $M$ rows of $N$
scatterers each, where the wave (or the random walker) is incident from left.
\ For this system, the amplitudes of transmission or reflection through
\textit{one} column become (instead of $t$ and $r$) the $M\times M$ matrices
$\widehat{\mathbf{T}_{1}}$ and $\widehat{\mathbf{R}_{1}}$ , whose element
$(\widehat{\mathbf{T}_{1}})_{ij}$ is the amplitude for a walker incident from
the left on row $i$ to exit to the right on row $j$, and an analogous
definition for $(\widehat{\mathbf{R}_{1}})_{ij}$. These amplitudes are
obtained by adding all the possible paths that lead from the input $i$ to the
output $j$ and are confined to \textit{one} single column of the medium. It
is, then, not difficult to prove that the expansion in loops (\ref{ldressedn})
remains valid as long as we replace in it $t$ and $r$ by these matrices, and
all the products are understood as products of matrices. Similarly, the
transmission $\widehat{\mathbf{T}_{N}}$ and the reflection $\widehat
{\mathbf{R}_{N}}$ through the strip of length $N$ are obtained through the
same replacement from the \ expressions (\ref{wavetn}) and (\ref{wavern}) \ respectively.

\ If the signal impinging on the system from the left is represented by an $M
$-component vector $\mathbf{I}$, the output vector $\mathbf{O}$ simply
becomes, for the case of transmission \
\begin{equation}
\mathbf{O}_{j}\,\,\mathbf{=\,}\sum_{i}\mathbf{I}_{i}(\widehat{\mathbf{T}_{N}%
})_{ij}\;,\label{osubj}%
\end{equation}
Since in this model the matrices $\widehat{\mathbf{T}_{1}}$ and $\widehat
{\mathbf{R}_{1}}$ have exactly the same symmetries, they will have the same
eigenvectors; those will also be the eignevectors of the complicated matrices
$\widehat{\mathbf{T}_{N}}$ and $\widehat{\mathbf{R}_{N}}$, since they are
expressed (through the loop expansion) as products of $\widehat{\mathbf{T}%
_{1}}$ and $\widehat{\mathbf{R}_{1}}$. The elements of the matrix
$\widehat{\mathbf{T}_{N}}$ \ involved in (\ref{osubj}) can be evaluated in
terms of those eigenvectors and the eigenvalues of $\widehat{\mathbf{T}_{1}}$
and $\widehat{\mathbf{R}_{1}}$ through standard algebraic techniques. The
problem is, then, essentially reduced to solving the eigenvalue problem of an
$M\times M$ matrix. \ %

\begin{figure}
[ptb]
\begin{center}
\includegraphics[
natheight=270.598694pt,
natwidth=354.418304pt,
height=175.1391pt,
width=228.9589pt
]%
{GQ6C7P04.wmf}%
\caption{Transmission through a system of length N=8 as a function of $t$, for
the cases of a 1-D chain, a strip of width $M=2$ (reflecting and absorbing
boundary conditions) and for a 2-D periodic system. The two-dimensional
systems correspond to the case $d=r$. The inset illustrates the
\textquotedblleft snaking\textquotedblright\ effect for $t\rightarrow0$.}%
\end{center}
\end{figure}
To conclude we will illustrate the treatment of the two dimensional problem by
considering the particularly simple case of periodic boundary conditions in
the lateral sides of the system (rows 1 and $M$). We will calculate the total
transmission through a strip of length $N$. Then,%

\begin{equation}
\mathbf{T}_{N}\equiv\sum_{j}\mathbf{O}_{j}=\sum_{i,\alpha,j}\mathbf{I}%
_{i}\;\lambda_{N}^{\alpha}\;\mathbf{E}_{i}^{\alpha}\mathbf{E}_{j}^{\alpha
}\label{formtn}%
\end{equation}
where the $\ \lambda_{N}^{\alpha}$ stand for the eigenvalues of the matrix
$\widehat{\mathbf{T}_{N}}$, and the $\mathbf{E}^{\alpha}$ for the
corresponding eigenvectors of the matrices $\widehat{\mathbf{T}_{1}}$ and
$\widehat{\mathbf{R}_{1}}$. \ From the symmetry of the problem readily follows
that one of the eigenvectors is $\mathbf{E}_{o}=\frac{1}{\sqrt{M}%
}(1,1,1,.....1)$ and all the others cancel when summed over $j$ in
(\ref{formtn}) by orthogonality, leading to $\mathbf{T}_{N}=\lambda_{N}%
^{o}\,\;(\sum_{i}\mathbf{I}_{i})$. The eigenvalue $\lambda_{N}^{o}\,\ $can be
simply obtained by replacing in the expansion (\ref{wavetn}) or its equivalent
for a PRW the coefficients $t$ and $r$ by the eigenvalues $\lambda_{1,T}%
^{o}\,\ $and $\lambda_{1,R}^{o}\,\ $(associated with $\mathbf{E}_{o}$) $\ $of
the matrices $\widehat{\mathbf{T}_{1}}$ and $\widehat{\mathbf{R}_{1}}$. Those
eigenvalues are simply $\sum_{j}(\widehat{\mathbf{T}_{1}})_{ij}$ for any row
$i$, and a similar expression for $\widehat{\mathbf{R}_{1}}$. For the case of
a PRW\cite{forwavelsewhere}, (where the walker has a probability $t$ of moving
forward , a probability $r$ of moving back, and a probability $d$ of turning
to either side, with $\ t+r+2d=1$), these eigenvalues can be obtained writing
all the possible paths starting from position $i$ that lead to transmission
(reflection) through \textit{any} row of a 1-column system:%

\[
\sum_{j}(\mathbf{T}_{1})_{ij}=t+2d^{2}(\frac{1}{1-t})(1+\frac{r}{1-t}+\left(
\frac{r}{1-t}\right)  ^{2}+.....)\equiv t+\frac{2\,d^{2}}{1-r-t}%
\]
The expansion in this expression corresponds to all paths with no reflection
in the vertical direction, one reflection in the vertical direction, etc.).
This is simply $\lambda_{1,T}^{o}\,=t+d$. Analogously we obtain $\lambda
_{1,R}^{o}\,=r+d$ . The total transmission for a system of length N becomes,
then:
\[
\mathbf{T}_{N}=(\sum_{i}\mathbf{I}_{i})\frac{(t+d)}{1+(N-1)(r+d)}%
\]
Figure 5 illustrates the transmission through a strip of length 8, for
periodic, absorbing and reflecting boundary conditions, and for a 1-D chain of
the same length, in the specific case $r=d$. The inset illustrates the effect
of \textquotedblleft snaking\textquotedblright, the existence of a non-zero
total transmission even in the case of a zero forward transmission $t$ at the
level of an individual scattering.

\begin{acknowledgement}
Charles Wizenfeld assisted with the numerical simulations. I benefited from
helpful discussions with Juanjo Saenz. The hospitality of the Departamento de
Fisica de la Materia Condensada, where the last part of this work was written
is gratefully acknowledged. Partially supported by the Summer Faculty
Fellowship of the Office of the Vice President of Academic Affairs of Yeshiva University.
\end{acknowledgement}


\begin{thebibliography}{99}                                                                                               %
\bibitem {Anderson}Anderson P W, \textit{Phys. Rev.} \textbf{109}, 1492 (1958)

\bibitem {gang of four}Abrahams E, Anderson P\ W, Licciardello D\ C and
Ramakrishnan T V, \ \textit{Phys. Rev. Lett.} \textbf{16}, 984 (1979)

\bibitem {Lagendijk}Van Albada M\ P and Lagendijk A, (1985), \textit{Phys.
Rev. Lett.} \textbf{55}, 2692.

\bibitem {Patrick Lee}Lee P\ A, in \textit{STATPHYS 16}, (H. E. Stanley, ed.),
North Holland (1986)

\bibitem {Van Albada y yo}Cwilich G and Fu Y, \textit{Phys. Rev.}
\textbf{B46}, 12015 (1992), \ Van Albada M\ P\ ,Van Tiggelen B et al,
\textit{Phys. Rev. Lett.} \textbf{66}, 3132 (1991)

\bibitem {Stone-Webb}Stone D, \textit{Phys. Rev. Lett} \textbf{54}, 2692
(1985), Webb R\ A\ ,Washburn S et al, \textit{Phys. Rev. Lett}. \textbf{54},
2696 (1985)

\bibitem {Genack}Stoytchev M and Genack A\ Z, \textit{Phys. Rev. Lett.}
\textbf{79}, 309 (1997), Garcia N and Genack A\ Z , \textit{Phys. Rev. Lett}
\textbf{63}, 1678 (1989)

\bibitem {Soukulis}Soukulis C in \textit{Diffuse Waves in Complex Media\ }(J.
P. Fouque ed.), 93-107, Kluwer (1998), Yablonovitch E and Leung K\ M,
\textit{Nature} \textbf{351}, 278 (1991)

\bibitem {Fink}Fink M, Time Reversed Acoustics, \textit{Phys. Today
}\textbf{50}, 34, March 1997.

\bibitem {Campillo}Campillo M et al. in \textit{Diffuse Waves in Complex
Media\ }(J. P. Fouque ed.), 383-404, Kluwer (1998)

\bibitem {Wiersma}Wiersma D, Van Albada M\ P\ and Lagendijk A
,\textit{\ Nature} \textbf{373}, 203 (1995), \ Beenakker C, \textit{Phys. Rev.
Lett.} \textbf{81}, 1829 (1998)

\bibitem {Wang}Wang L et al., \textit{Science} \textbf{253}, 769 (1991), and
also \textit{Appl. Optics} \textbf{32,} 5043 (1993).

\bibitem {Kaiser}Kaiser R, in \textit{Diffuse Waves in Complex Media\ }(J. P.
Fouque ed.), 249-288, Kluwer (1998)

\bibitem {Wiersma2}Wiersma D, Bartolini P and Lagendijk A, \textit{Nature}
\textbf{390}, 691 (1997)

\bibitem {Kogan}Kogan E and Kaveh M, \textit{Phys. Rev}. \textbf{B48}, 9404
(1993) and \textit{B52}, 3813 (1995), \ Kogan E et al. \textit{Physica}
\textbf{A200}, 469 (1993).

\bibitem {yo-OSA}Cwilich G in \textit{OSA Proceedings on Advances in Optical
Imaging and Photon Migration} (R. Alfano, ed.),7-11, Opt. Soc. Of America (1994)

\bibitem {Sebbah}Sebbah P, Legrand O and Genack A\ Z, \textit{Phys. Rev.
}\textbf{E59, }2406 (1999)

\bibitem {Feller}Feller W, \textit{An Introd. to Probability Theory and its
Applications}, \ John Wiley (1968)

\bibitem {weissrubin}Weiss G\ H and Rubin R\ J, \textit{\ Adv. in Chemical
Physics} \textbf{52}, 363 (1983).

\bibitem {weiss}Weiss G\ H, \textit{J. Stat. Phys.} \textbf{15}, 157 (1976)

\bibitem {martinez}Martines A\ S, Doctoral Thesis, Universit\'{e} Joseph
Fourier (1995)

\bibitem {ed-kav-shap}Edrei I, Kaveh M and Shapiro B, \textit{Phys. Rev.
Lett.} \textbf{62}, 2120 (1989)

\bibitem {seb2}Vanneste C\ P, Sebbah P and Sornette D, \textit{Europhysics
Letters \ }\textbf{17}, 715 (1992)

\bibitem {seb3}Sebbah P, Doctoral Thesis, Universit\'{e} de Nice-Sophia
Antipolis (1993)

\bibitem {stepabra}Abrahams E and Stephen M\ J, \textit{J. Phys.}
\textbf{C13}, L377 (1980)

\bibitem {andereck}Andereck B\ S, Abrahams E, \textit{J. Phys.} \textbf{C13},
L383, 1980

\bibitem {cwilklein}Unpublished

\bibitem {weitz}Pine D\ J\ , Weitz D\ A, Maret G, Wolf P\ E, Herbolzheimer E
and Chaikin P\ M in \textit{Scattering and Localization of classical waves in
random media} (Ping \ Sheng ed.),312-372, World Scientific, (1990).

\bibitem {forwavelsewhere}The analogous results for the amplitudes of a wave
will be published elsewhere.
\end{thebibliography}
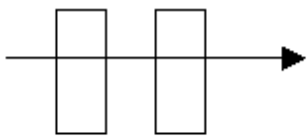
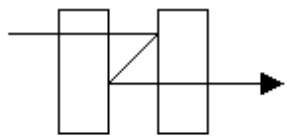
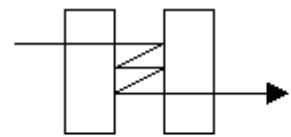
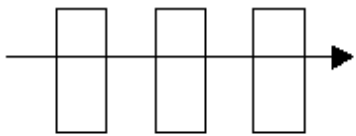
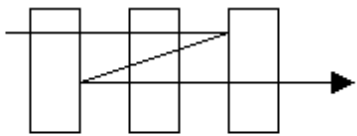
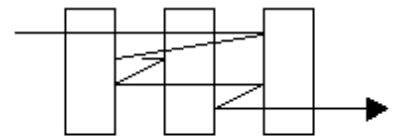
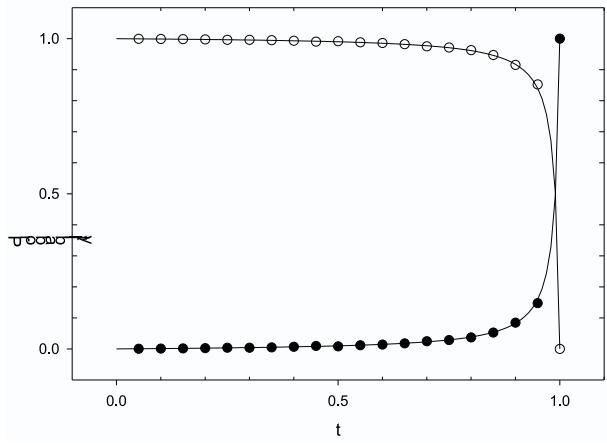
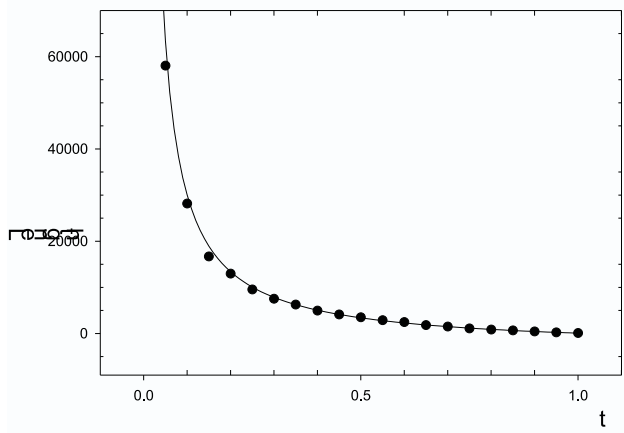
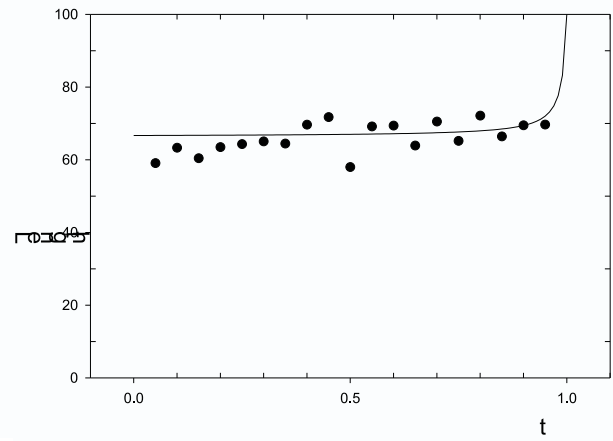
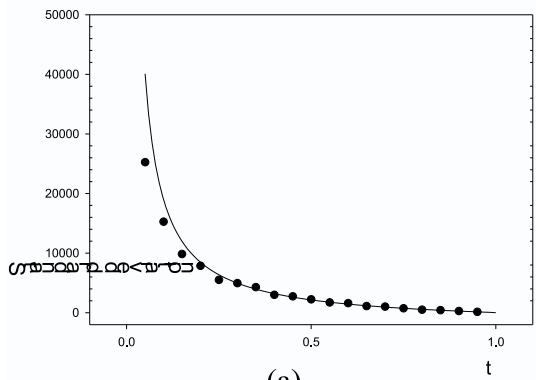
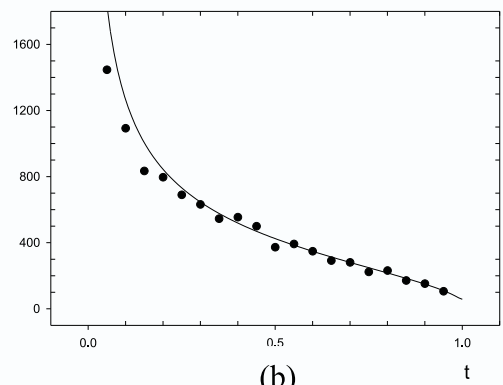
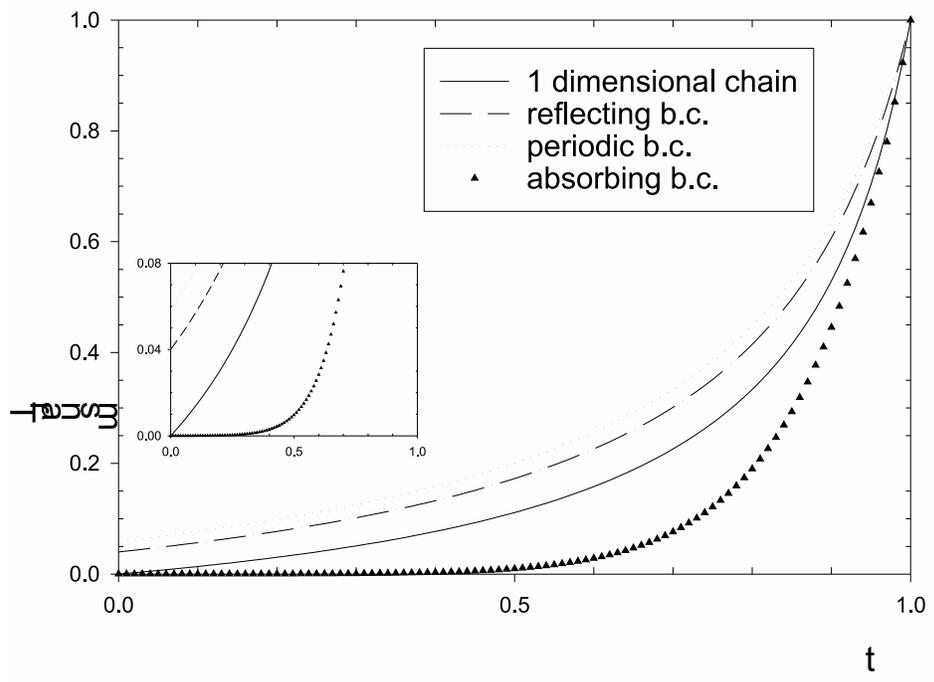
\end{document}